\documentclass{elsart}
\usepackage{amssymb}
\usepackage[dvips]{graphicx}% Include figure files

\begin{document}

\begin{frontmatter}


\title{Interparticle interaction and structure of deposits for
competitive model in (2+1)- dimensions}
\author[Univ]{L.A. Bulavin}
\author[IBCC]{N.I. Lebovka\corauthref{cor}}
\corauth[cor]{Corresponding author}
\ead{lebovka@mail.kar.net}
\author[KPI]{V.Yu. Starchenko}
\author[IBCC]{N.V. Vygornitskii}
\address[Univ]{Department of Physics, Taras Shevchenko National University, 6, pr. Glushkova, Kyiv 03127, UA}
\address[IBCC]{F.D. Ovcharenko Biocolloid Chemistry Institute NASU, 42, blvd. Vernadskogo, Kyiv 03142, UA}
\address[KPI]{Physico-Technical Institute, National Technical University of Ukraine KPI, 37, prosp. Peremogy, Kyiv 03056, UA}

\begin{abstract}
% Text of abstract
A competitive (2+1)-dimensional  model of deposit formation, based
on the combination of random sequential absorption deposition
(RSAD), ballistic deposition (BD) and random deposition (RD)
models, is proposed. This model was named as
RSAD$_{1-s}$(RD$_f$BD$_{1-f}$)$_s$. It allows to consider
different cases of interparticle interactions from complete
repulsion between near-neighbors in the RSAD model ($s=0$) to
sticking interactions in the BD model ($s=1, f=0$) or absence of
interactions in the RD model ($s=1$, $f=0$). The ideal
checkerboard ordered structure was observed for the pure RSAD
model ($s=0$) in the limit of $h \to \infty$. Defects in the
ordered structure were observed at small $h$. The density of
deposit $p$ versus system size $L$ dependencies were investigated
and the scaling parameters and values of $p_\infty=p(L=\infty)$
were determined. Dependencies of $p$ versus parameters of the
competitive model $s$ and $f$ were studied. We observed the
anomalous behaviour of the deposit density $p_\infty$ with change
of the inter-particle repulsion, which goes through minimum on
change of the parameter $s$. For pure RSAD model, the
concentration of defects decreases with $h$ increase in accordance
with the critical law $\rho\propto h^{-\chi_{RSAD}}$, where
$\chi_{RSAD} \approx 0.119 \pm 0.04$.
\end{abstract}

\end{frontmatter}

\newpage

\section{Introduction}

Recently the study of the nonequilibrium processes of deposit
formation became one of the most active research areas in physics
and chemistry because of both fundamental and practical interest
\cite {Family91}. The essential progress in this field was reached
owing to application of computer simulations methods for studying
the problems of adsorption, formation of thin films and coating.
There exists a limited number of computer models, allowing to
carry out effective calculations and to establish the main scaling
laws for formation of deposits \cite {Barabasi95}.

Among the most popular models for deposit formation simulation are
the models of random sequential adsorption (RSA), ballistic
deposition (BD) and random deposition (RD)
\cite{Barabasi95,Evans93}. In these models, the particles get
fixed after deposition and don't move, so these models describe
the growth processes far from equilibrium. It is also assumed that
particles are rigid and can not overlap. In RSA model, particles
get fixed at some distance one from another
\cite{Evans93,Privman95}. In RSA model with the nearest neighbor
(NN) exclusion, configurations with the particles having nearest
neighbors are eliminated. In fact, it means existence of some
repulsion between particles. In RSA model, all sites in the
lattice can be filled with equal probability. However, this
requirement is not fulfilled on deposit formation. In this case,
the previously deposited particles can screen free sites, located
below. So, we named the variant of RSA model for deposit formation
as RSA deposition model (RSAD). In BD model, the particles stick
at the point of their first contact. It means existence of the
short-range attraction. In RD model, the particles deposit without
sticking and it means existence of the short-range repulsion.
Recently, a number of mixed or competitive models were proposed.
They are based on consideration of deposition for different kinds
of particles \cite{Mixed1,Mixed2,Mixed3}

In this work, the competitive (2+1)-dimensional model with three
kinds of particles, depositing according to the rules of random
sequential absorption deposition (RSAD), ballistic deposition (BD)
and random deposition (RD), is investigated. This model allows to
explore a wide class of systems with different interparticle
interactions, which vary from a complete repulsion as for RSAD
model to short-range attraction as for BD model.

The paper is organized as follow. The model is described in
section \ref{model}. In section \ref{results}, the scaling
behaviour of the deposit density and concentration of the
structure defects are discussed for different values of the
competitive model parameters. Concluding remarks are presented in
section \ref{conclusions}

\section{Model}
\label{model}

The competitive (2+1)-dimensional model, which combines models of
the random sequential adsorption deposition (RSAD), ballistic
deposition (BD) and random deposition (RD), named by as RSAD
$_{1-s}$ (RD$_f$BD$_{1-f}$)$_s $, where $s$ and $ f$ parameters
are fractions of particles with different kinds of interparticle
short-range interaction potentials. A particle of RSAD or BD or RD
kind randomly falls straight down onto a growing surface, one at a
time, and deposits at a site of the cubic lattice.

Parameter $s$ characterizes the short-ranged next-near-neighbor
repulsion of particles in the deposit, and parameter $f$
characterizes adhesion of particles. In the extreme case of $s =
0$ all the particles are of RSAD kind. This case corresponds to
the strong repulsion between particles  in NN sites and they can
deposit only in the next NN sites. When $s = 1 $, $p = 0$, all the
particle are of BD kind and newcomer sticks to the deposits at a
point of its first contact. When $s = 1$, $ p = 1$, all the
particle are of RD kind. This case corresponds to existence of the
short-ranged repulsion between particles and formation of
completely compact deposits.

The density of a deposit may be defined as
\begin{equation} \label{e2}
p={N}/{V}
\end{equation}
where N is the number of deposited particles, and $V$ is the
volume of deposit.

The value of $p$ is calculated at a moment when saturation was
reached in the system with the volume $V=L\times L\times L$. At
this moment, filling of the system with new particles was
terminated.

%======================================== Fig.1 ========================================
\begin{figure}[top]
\begin{center}
\includegraphics[angle=0,width=14.0cm,clip=]{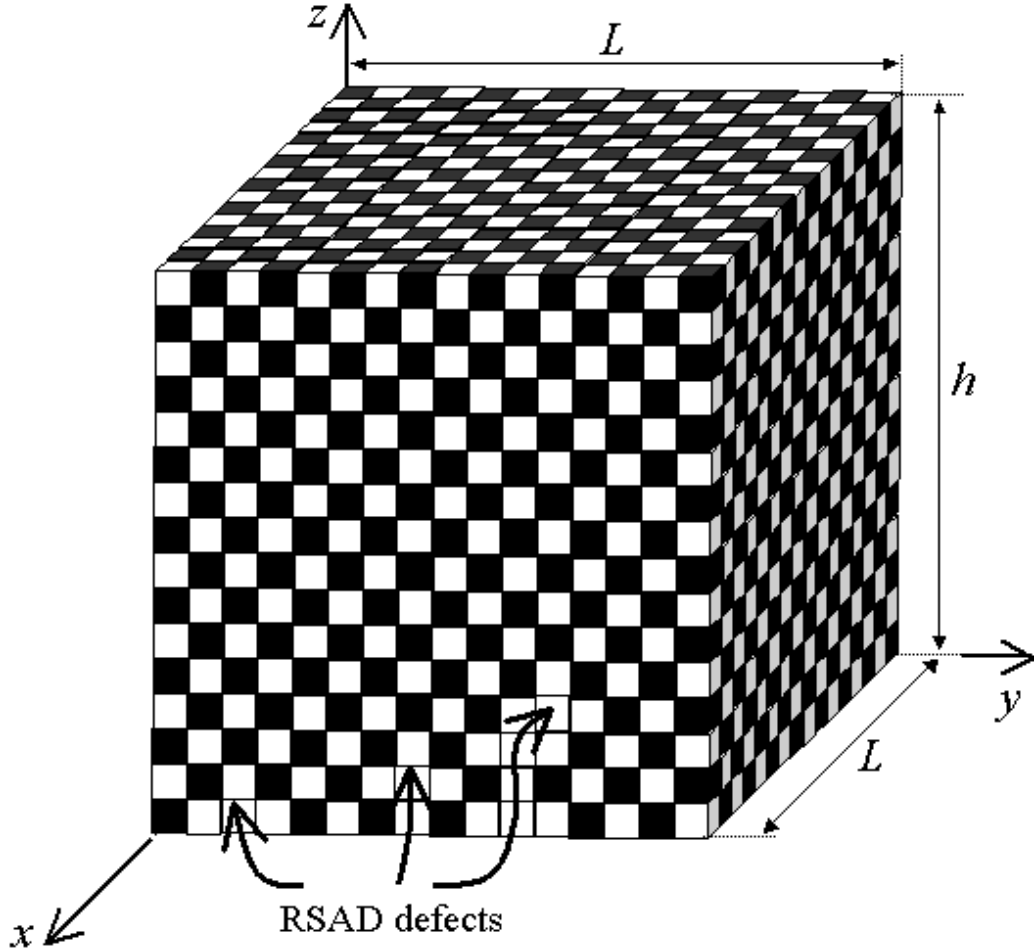}
\end{center}
\caption{ Schematic presentation of the deposit for pure RSAD
model ($s = 0$). Black cubes correspond to the sites filled with
particles. In the ideal checkerboard-ordered structure each
particle is surrounded by 6 empty sites. The examples of defects
are also shown.} \label{f01}
\end{figure}
%---------------------------------------------------------------------------

For RSAD model, the structure of deposit becomes more regular at
large deposit height $h$, and in the extreme case of
 $h\to \infty$,  the ideal checkerboard-ordered structure get formed.
In this ideal structure, the empty lattice sites alternate with
occupied ones and each empty (or occupied) site has exactly $z=6$
occupied (or empty) NN sites. But at small values of $h$ there
exist some defects in the regular structure (Fig. \ref{f01}), when
NN exclusion number $z$ is different from 6.

The concentration of defects $\rho(h)$ may be defined as
\begin{equation}
\label{e3} \rho(h)=1-\bar{z}(h)/6,
\end{equation}
where $\bar{z}(h)$ is the NN exclusion number averaged over all
the occupied and empty sites in the horizontal layer with the
constant $h$.

The periodic boundary conditions are applied along the horizontal
directions $x$ and $y$. The size of the base $L$ was varied within
the interval of $L=10-400$ and deposit height was $h \leq 200L$.
The values of $p$ and $\rho(h) $ were averaged over $100-1000$ of
different configurations for each fixed set of parameters $ s $
and $ f $.

\section{Results and discussion}
\label{results}

At the beginning we studied the scaling behavior of the deposit
density $p$ in cubes with varying sizes $L\times L\times L$. A
clear scaling law of the following  type was observed:
\begin{equation}
p=p_\infty + a L ^{-1/\nu}, \label{scal}
\end{equation}
Here, $p_\infty=p(L \to \infty)$, $\nu$  is the scaling exponent
and $a$ is the amplitude.

%======================================== Fig.2 ========================================
\begin{figure}[top]
\begin{center}
\includegraphics[angle=0,width=14.0cm,clip=]{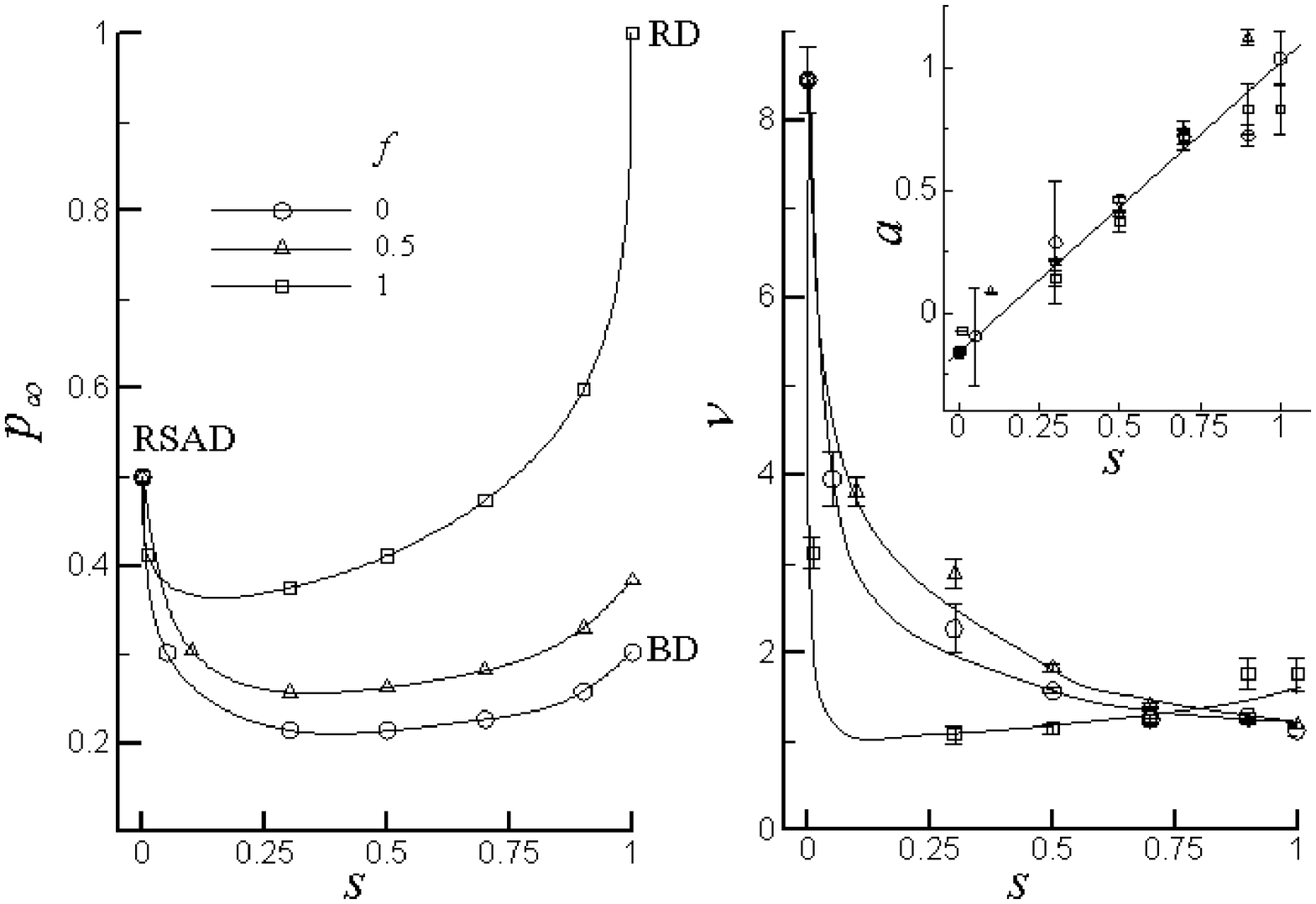}
\end{center}
\caption{Density of deposit $p$ and  scaling exponent $\nu$ versus
$s$ in equation \ref{scal} at different values of $f$. Insert
shows similar plots of scaling amplitude $a_p$ versus $s$.}
\label{f02}
\end{figure}
%---------------------------------------------------------------------------
Figure \ref{f02} shows $p_\infty$, $\nu$ and $a$ versus $s$
dependencies at different values of $f$. As we have noted
previously, for RSAD model ($s=0$)  in the limits of $L \to
\infty$ and  $h \to \infty$ the ideal checkerboard ordered
structure get formed, that corresponds to $p_\infty=0.5$. With
increasing of $s$ when $f$ is a constant the density of a deposit
$p_\infty$ always goes through the minimum at $s=s_{\min}$. The
value of $s_{\min}$ increases with decreasing of $f$. The point of
$s=1$ and $f=0$ corresponds to a pure BD model for which
$p_\infty=0.3000$ (this value in accordance with data
\cite{Krug91}). The point of $s=1$ and $f=1$ corresponds to the
pure RD model, when $p_\infty=1$ and compact deposit forms.

At first glance, the $p_\infty$ versus $s$  behavior is somewhat
anomalous. Really, the effective repulsion between particles
decreases with increase of $s$ (at constant $f$). So, we should
expect compacting of deposit and increase of $p_\infty$. Another
interesting feature is that the scaling exponent $\nu$ decreases
abruptly with $s$ decrease in the interval of $s<0.1$. In fact, we
observe two different scaling regimes  with large and small values
of scaling exponent $\nu$. For the pure RSAD model ($s=0$),
scaling exponent is $\nu_{RSAD}=8.4 \pm 0.3$. In a regime far from
pure RSAD model, the scaling exponent falls within the interval of
$\nu=1.0-1.8$. The amplitude $a$ continuously grows with $s$
increase and it approaches zero at $s<0.1$, which results in large
errors of  $\nu$ estimation (see insert in Fig. \ref{f02}).

The nature of unusual behavior of the deposit density $p_\infty$
with variation of $s$ may be understood more clearly from the
analysis of defects evolution during the deposit growth. The
initial abrupt decreasing of the deposit density with $s$ increase
(at $s\lesssim 0.1$) reflects generation of defects in the ideal
checkerboard-ordered structure. Naturally this results in
loosening of the  structure of deposit. The point of density
minimum at $s=s_{min}$ corresponds to a kind of equilibrium
between the processes of birth and vanishing of defects. With
further $s $ increase, effectiveness of regeneration of the ideal
checkerboard-ordered structure decreases. Finally, it results in
increase of $p_\infty$ with $s$ increase.

%======================================== Fig.3 ========================================
\begin{figure}[top]
\begin{center}
\includegraphics[angle=0,width=14.0cm,clip=]{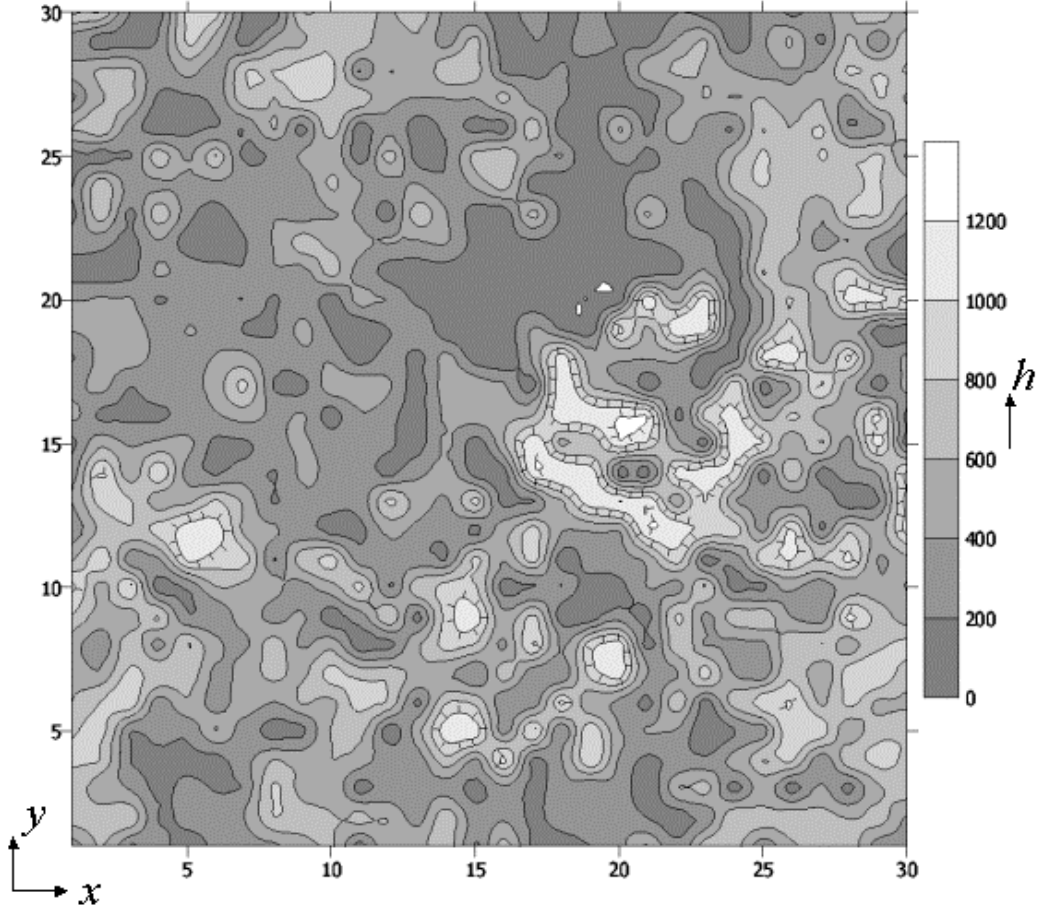}
\end{center}
\caption{ A typical map of the heights of defect annihilation in
the plane of $x-y$. For clearer presentation, a small system of
the size $30\times 30\times 1400 $ was taken. } \label{f03}
\end{figure}
%---------------------------------------------------------------------------

The described behavior of $p_\infty$ can be compared with behavior
of the concentration of defects $ \rho(z) $. Figure \ref {f03}
shows the typical map of the defect annihilation heights in the
plane of $x-y$ for the pure RSAD model. Here, the lighter color
corresponds to the higher deposit coordinate $z$, at which defects
disappear in the ideal RSAD structure.

At initial moment of deposit formation, there arise a lot of
defects in the ideal RSAD structure. This is a result of the
absence of correlations between deposition in the columns at small
height of the deposit. These defects can spread in the direction
of deposit formation along $ z $ axis and their concentration in
each layer $ \rho (z) $ increases or decreases owing to the
processes of defect birth (at $s\neq 0$)  or annihilation,
respectively.

%======================================== Fig.4 ========================================
\begin{figure}[top]
\begin{center}
\includegraphics[angle=0,width=14.0cm,clip=]{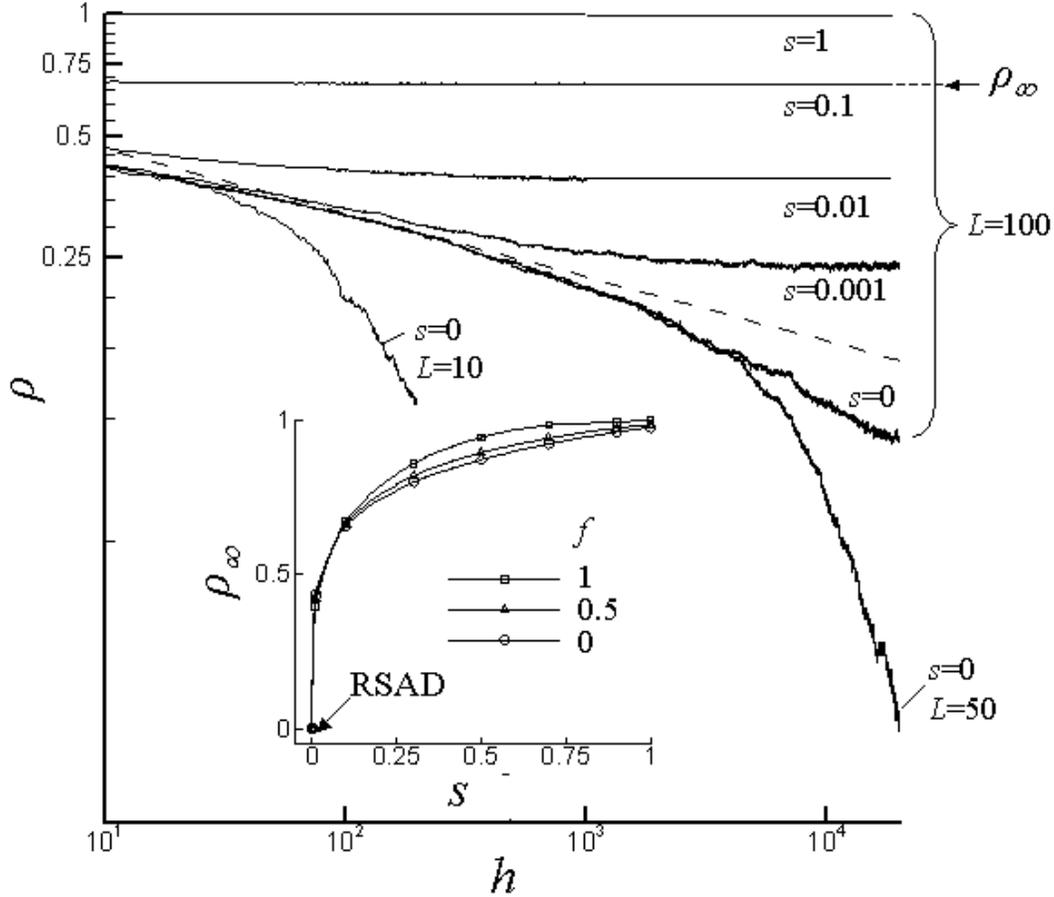}
\end{center}
\caption{Concentration of RSAD defects $\rho$ versus the deposit
height $h$ at $f=1$ and different $s$ and $L$. The dashed line
corresponds to the power equation (\ref{nheq}) with the slope
$\chi_{RSAD} \approx 0.119 \pm 0.04$. The value of $\rho_\infty$
corresponds to the limit $h\rightarrow \infty$. Insert shows the
limiting values of concentration  of the RSAD defects
$\rho_{\infty}$ (at $h \to \infty$) and density of a deposit
$p_\infty$ (at $h=L \to \infty$) versus $s$ at different values of
$f$. } \label{f04}
\end{figure}
%---------------------------------------------------------------------------

Figure \ref{f04} shows the examples of defects concentration
$\rho$ versus deposit height $h$ dependences for the case when
$f=1$. The similar dependences were also observed  for other
values of $f$.

For pure RSAD model ($s = 0 $) the concentration of defects $ \rho
$ continuously decreases with increase of the deposit height $h$.
In the limit of $ h\to\infty$ the defectless checkerboard-ordered
structure get formed. In the latter case, the obvious scaling
behavior of $\rho (h)$ dependencies is observed for systems with
different size of base $L$. We believe that a simple power law can
be applied for description of $\rho (h)$ behavior in the limit of
$L\rightarrow \infty$
\begin{equation}
 \rho \propto h^{-\chi_{RSAD}} %\mbox{ at } L\rightarrow \infty
\label{nheq}
\end{equation}
where $\chi_{RSAD}$ is the  scaling exponent of defects
annihilation for the pure RSAD model.

This behaviour is rather similar to the critical annihilation of
inter-domain boundaries in the model of competitive growth by
Saito and Muller-Krumbhaar (SMK) \cite{SMK} ( see also,
\cite{Vandewalle,Lebovka}).

Prediction of the scaling exponent in critical annihilation is not
trivial. For normal Brownian motion of annihilating defects the
scaling exponent should be $\chi_{B}=1/2$ \cite{Spouge88}. In
$1+1$ dimensional SMK-model another scaling exponent $\chi_{SMK}
\approx 2/3$ \cite{SMK} was observed. This result was explained by
existence of different mechanisms of the critical annihilation of
defects in SMK and random walk motion models. It was conjectured
that the lateral displacements of inter-domain boundaries in  the
SMK model are controlled by the same processes as those causing
roughening the outer interface width and $\chi_{SMK} \approx
\alpha$, where $\alpha$ is the roughening exponent. On the other
hand, the value of $\alpha$ can reflect the growth mechanism,
geometrical confinement of growing interface, space dimension,
etc.\cite{Derrida91}. For example, for usual models of KPZ class
of universality \cite{KPZ} the roughening exponent is $\alpha=1/2$
in $1+1$ dimension and is $\approx 0.2-0.4$ in $2+1$ dimension
\cite{Barabasi95}.

For pure RSAD model ($s = 0 $) and an isotropic system ($L \times
L \times L$), we can estimate the value of the scaling exponent in
Eq. \ref{nheq} as $\chi_{RSAD}=1/\nu_{RSAD}\approx 0.119 \pm
0.04$, where $\nu_{RSAD}=8.4 \pm 0.3$ is the scaling exponent for
the density of deposit in Eq. \ref{scal}.  Indeed, we believe that
the scaling changes in the deposit density $p$ and in the
concentration of defects $\rho$ are controlled by the same
process, and $p(L)\propto \rho(L)$. The dashed line in Fig.
\ref{f04} corresponds to the slope, estimated from scaling of the
deposit density. But for anisotropic systems $L \times L \times h$
 ($h>L$) , the concentration of defects decrease more quickly than
it is predicted by the power law in Eq. \ref{nheq}, and the most
obvious deviations are observed with increase of the system
anisotropy degree at high $h$ (see Fig.\ref{f04}).

In a general case of the competitive  model RSAD $_{1-s}$
(RD$_f$BD$_{1-f}$)$_s$, the patterns of defect distributions in a
deposit can be more complicated. Here, defects can birth or can vanish
and competition between these processes results in formation of a
certain defected structure with the finite concentration of
defects. For mixed model at $s\neq0$, the critical behavior of
(\ref{nheq}) type disappears. In the limit of indefinitely large
$h$ ($h\rightarrow \infty$), the concentration of defects remain
finite and has a nonzero value $\rho_\infty$. With $s$ increase,
which corresponds to the weakening of repulsion between the
particles, increase of $\rho_\infty$ is observed (See insert in
Fig.\ref{f04}), and in the limit of pure RD model, ($s=1$, $f=1$),
$\rho_\infty=1$, because deposit is compact in RD model and even
local checkerboard-ordered structure is absent.

\section{Conclusions}
\label{conclusions}

It was demonstrated that implementation of the RSA rules for 3d-
growth of deposit on a cubic lattice with the nearest-neighbor
exclusions results in formation of a checkerboard-ordered
structure. The initial disorder of this structure critically
disappears when the deposit grows and, finally, an ideal
checkerboard-ordered structure gets formed. In a competitive
RSAD$_{1-s}$(RD$_f$BD$_{1-f}$)$_s$ model, the deposit density goes
through the minimum with variation of $s$ parameter at constant
$f$. Such behavior reflects the influence of competitive processes
related with the birth and annihilation of defects.

%\acknowledgements
\section*{Acknowledgements}
This work was supported in part by the NASU under the program
"Nanosystems, Nanomaterials and Nanotechnologies" and Grants No.
2.16.1.4 (0102V007058) and 2.16.2.1(0102V007048).

\end{document}